\documentclass[aps,twocolumn,preprintnumbers,amsmath,amssymb,superscriptaddress]{revtex4}
\usepackage{epsfig}
\usepackage{graphicx}
\usepackage{dcolumn}
\usepackage{bm}
\usepackage{float}
\usepackage{hyphenat}
\usepackage[dvipsnames]{xcolor}
\usepackage{physics}
\usepackage[T1]{fontenc}
\usepackage{gensymb}
\usepackage[utf8]{inputenc}
\usepackage{amsmath}
\usepackage{amsthm}

\DeclareUnicodeCharacter{2212}{\ensuremath{-}}

\newcommand{\heriotwatt}{Institute of Photonics and Quantum Sciences, SUPA, Heriot-Watt University, Edinburgh EH14 4AS, UK}
\newcommand{\TsukubaKenji}{Research Center for Functional Materials, National Institute for Materials Science, 1-1 Namiki, Tsukuba 305-0044, Japan}
\newcommand{\TsukubaTakashi}{International Center for Materials Nanoarchitectonics, National Institute for Materials Science,  1-1 Namiki, Tsukuba 305-0044, Japan}
\newcommand{\insa}{Université de Toulouse, INSA-CNRS-UPS, LPCNO, 135 Avenue de Rangueil, 31077 Toulouse, France}
\newcommand{\tud}{Institute of Condensed Matter Physics,
Technische Universit\"at Darmstadt, 64289 Darmstadt, Germany}

\begin{document}

\title{Highly Tunable Ground and Excited State Excitonic Dipoles in Multilayer 2H-MoSe$_2$}

\author{Shun Feng}
\thanks{These two authors contributed equally}
\author{Aidan Campbell}
\thanks{These two authors contributed equally}
\affiliation{\heriotwatt}
\author{Mauro Brotons-Gisbert}
\email{m.brotons_i_gisbert@hw.ac.uk}
\affiliation{\heriotwatt} 
\author{Daniel Andres-Penares}
\affiliation{\heriotwatt} 
\author{Hyeonjun Baek}
\affiliation{\heriotwatt}
\author{Takashi Taniguchi}
\affiliation{\TsukubaTakashi}
\author{Kenji Watanabe}
\affiliation{\TsukubaKenji}
\author{Bernhard Urbaszek}
\affiliation{\tud}	
\author{Iann C. Gerber}
\affiliation{\insa}
 \author{Brian D. Gerardot}
 \email{B.D.Gerardot@hw.ac.uk}
 \affiliation{\heriotwatt}
    
\date{\today}

\begin{abstract}
The fundamental properties of an exciton are determined by the spin, valley, energy, and spatial wavefunctions of the Coulomb bound electron and hole. In van der Waals materials, these attributes can be widely engineered through layer stacking configuration to create highly tunable interlayer excitons with static out-of-plane electric dipoles, at the expense of the strength of the oscillating in-plane dipole responsible for light-matter coupling. Here we show that interlayer excitons in bi- and tri-layer 2H-MoSe$_2$ crystals exhibit electric-field-driven coupling with the ground ($1s$) and excited states ($2s$) of the intralayer A excitons. We demonstrate that the hybrid states of these distinct exciton species provide strong oscillator strength, large permanent dipoles (up to $0.73 \pm 0.01$ enm), high energy tunability (up to $\sim$ 200 meV), and full control of the spin and valley characteristics such that the exciton g-factor can be manipulated over a large range (from -4 to +14). Further, we observe the bi- and tri-layer excited state ($2s$) interlayer excitons and their coupling with the intralayer excitons states ($1s$ and $2s$). Our results, in good agreement with a coupled oscillator model with spin (layer)-selectivity and beyond standard density functional theory calculations, promote multilayer 2H-MoSe$_2$ as a highly tunable platform to explore exciton-exciton interactions with strong light-matter interactions.

\end{abstract}


\maketitle

\section{Introduction}
A range of exotic collective effects are predicted to arise from dipolar interactions \cite{butov2004condensation, lahaye2009physics}, which have a quadratic dependence on the magnitude of the static electric dipoles (\emph{p}). For example, strong dipolar interactions may result in exciton crystals, which exhibit ordering due to a balance between exciton kinetic energy and many-body Coulomb interactions \cite{PhysRevLett.98.060405, PhysRevB.75.134302, PhysRevLett.125.255301, astrakharchik2021quantum, bondarev2021crystal}, or lead to nonlinear exciton switches which can reach the quantum limit when the strength of the interaction is larger than the exciton’s radiative linewidth \cite{chang2014quantum,BUTOV20172,delteil2019towards}. Hence, in the solid-state, much emphasis has been placed on engineering interlayer excitons with large \emph{p} in pioneering III-V heterostructures \cite{high2012spontaneous} and more recently in transition metal dichalcogenide (TMD) heterostructures \cite{fogler2014high, erkensten2021exciton, PhysRevB.103.L041406} which host excitons with huge binding energies and thus small Bohr radii \cite{chernikov2014exciton, PhysRevLett.113.026803} that enable high exciton densities \cite{wang2019evidence, PhysRevLett.120.207401}. In TMD heterostructure devices, tunable interlayer excitons with large $p$ have been realized in homobilayers \cite{arora2017interlayer, arora2018valley, peimyoo2021electrical, lorchat2021excitons, leisgang2020giant, horng2018observation, wang2018electrical, kipczak2022analogy, paradisanos2020controlling, sung2020broken,shimazaki2020strongly} and heterobilayers \cite{rivera2015observation, ciarrocchi2019polarization, jauregui2019electrical, tang2021tuning}, even at the single exciton level \cite{kremser2020discrete, baek2020highly, li2020dipolar}. However, many of the exotic collective effects underpinned by strong dipolar interactions remain to be observed, motivating further exploration of interlayer excitons and ways to manipulate their spin and optical properties. For example, it is desirable to increase the electron-hole spatial separation beyond the interlayer distance, but not at the cost of vanishing oscillator strength. This goal is intrinsically difficult for bare interlayer exciton states (e.g. in TMD heterobilayers with Type II interfaces). Recent experimental and theoretical efforts \cite{leisgang2020giant, peimyoo2021electrical, PhysRevB.99.035443} suggest that each `bare exciton' is actually composed of a mixture of other exciton wave functions, and their mixture can be tuned by detuning energy between transitions. Here we use the term bare exciton to denote the majority of each transition for simplicity. While previous studies focused on hybrid interlayer excitons formed in adjacent `natural' homobilayers \cite{peimyoo2021electrical, lorchat2021excitons, leisgang2020giant}, an open question remains if larger hybrid interlayer dipoles, with the electron and hole highly confined in separate layers, can be generally obtained in multilayer TMD platforms. Further, due to the decreasing oscillator strengths of excited Rydberg-like states \cite{chernikov2014exciton, PhysRevLett.113.026803, stier2018magnetooptics}, the observation of excited states of interlayer excitons has proven elusive to date. Characterisation of the excited state spectrum of interlayer excitons provides additional information about their basic properties and a potential means to further engineer dipolar interactions by taking advantage of their larger Bohr radii, which has been crucial to realize optical nonlinearity \cite{walther2018giant} and excitonic analogues of spatially ordered structures in ultracold atomic gases  \cite{schauss2012observation}. To address these issues, we take advantage of hybridised excitons in bilayer (2L) and trilayer (3L) 2H-MoSe$_2$ composed of the bare exciton states of the interlayer excitons (both $1s$ and $2s$) and the ground ($1s$) and excited ($2s$) states of the intralayer A-excitons. We simultaneously demonstrate strong oscillator strength and wide tunability of the fundamental properties (spin, valley, energy, and spatial wavefunction) of the hybrid exciton species in multilayer MoSe$_2$.

\begin{figure*}
    \begin{center}
    	\includegraphics[scale= 0.14]{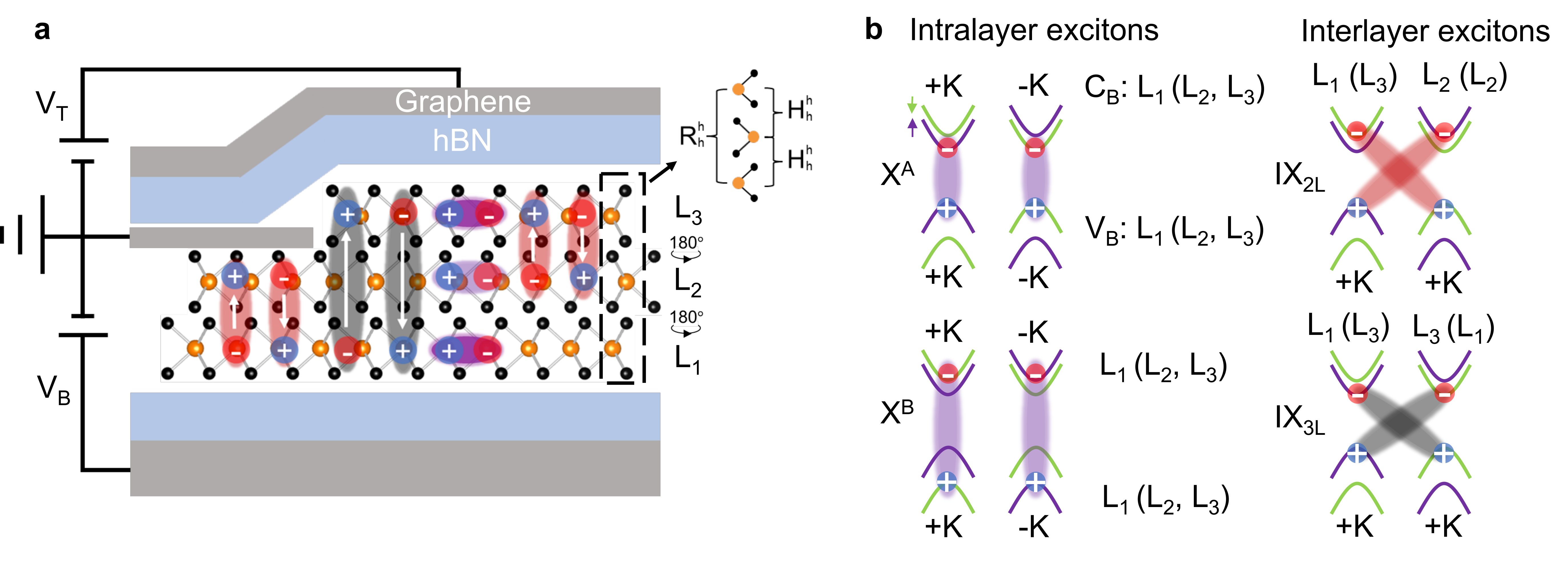}
    \end{center}
    \caption{\textbf{Intralayer and interlayer excitons in a terraced 2L/3L 2H-MoSe$_2$ sample}. \textbf{a}, Sketch of the sample and device used in this work: a 2H-MoSe$_2$ crystal with 2L- and 3L-thick terraces in a dual-gated device configuration. Electrons and holes (red and blue circles, respectively) can be localised either in the bottom (L$_1$), middle (L$_2$), or top (L$_3$) MoSe$_2$ layers, giving rise to different species of strongly bound intralayer and interlayer excitons. The vertical white arrows indicate the direction of the permanent electric dipoles for the different interlayer excitons. The interlayer stackings are highlighted as $R^h_h$ between L$_1$ and L$_3$ and $H^h_h$ between L$_1$ (L$_2$) and L$_2$ (L$_3$), where $h$ denotes the hexagon center in the crystal lattice of each layer. \textbf{b}, Spin–valley configurations of the intralayer and interlayer excitons depicted in panel a: intralayer A and B excitons (purple shaded ovals), and bilayer and trilayer interlayer excitons (red and grey shaded ovals, respectively). The layer labels (L$_1$-L$_3$) on each panel indicate the layer origin of the corresponding electronic states in the conduction (C$_B$) and valence (V$_B$) bands. The purple (green) bands correspond to spin up (down).}
    \label{fig1}
\end{figure*}


\section{Results and discussion}
\subsection{Device structure and introduction to excitons in multilayer 2H-MoSe$_2$}

Fig. \ref{fig1}a shows a sketch of our device, which consists of a terraced 2H-MoSe$_2$ flake with bilayer (2L) and trilayer (3L) regions encapsulated by hexagonal boron nitride (h-BN) layers with nearly identical thicknesses ($\sim$18 nm). Graphene layers act as electrical contacts for the MoSe$_2$ crystal and the top and bottom gates (see Suppl. Note S1 for further details of the device fabrication). In our experiments, the MoSe$_2$ contact is grounded while we apply voltages to the top and bottom gates, labeled as V$_T$ and V$_B$, respectively. This configuration allows us to apply a vertical electric field with a magnitude V$_E$ (where V$_E$ = V$_B$ = -V$_T$) while keeping the carrier concentration in the MoSe$_2$ sample constant at charge neutrality. Similar to other TMDs, each layer in our terraced 2H-MoSe$_2$ flake hosts tightly bound intralayer excitons with both ground ($1s$) and excited ($2s$, $3s$, etc) states \cite{chernikov2014exciton}. Ground and excited intralayer exciton states present the same exact spin-valley configuration. In the case of the lowest energy excitons in MoSe$_2$ (the so called A excitons, X$^A$), both the ground and excited exciton states originate from Coulomb-bound electron-hole pairs in the lower- (higher-) lying conduction (valence) band at $\pm$K (see Fig. \ref{fig1}b), respectively, which endows them with the same optical selection rules: excitons at $\pm$K couple to $\sigma^{\pm}$-polarised light \cite{xiao2012coupled}. In addition to X$^A$, there are intralayer B excitons (X$^B$) composed of the electron (hole) at the top (bottom) of the conduction (valence) band at $\pm$K respectively, with a considerable energy difference and opposite spin index compared to X$^A$.

Beyond intralayer exciton states, multilayer TMDs also host interlayer excitons, in which the layer-localised electron wave functions can  bind to holes with wave functions confined predominantly within adjacent layers or spread across several layers, giving rise to excitons with spatially extended wave functions \cite{arora2017interlayer,wang2018electrical,horng2018observation,leisgang2020giant,lorchat2021excitons,peimyoo2021electrical}. Interlayer and intralayer excitons have been shown to coexist in multilayer \cite{arora2018valley} and bilayer 2H-MoSe$_2$ \cite{horng2018observation}. Figure \ref{fig1}a shows a sketch of the possible intra- and interlayer exciton spatial configurations in our terraced 2L/3L 2H-MoSe$_2$, where we have assumed that, despite the possible spatial spread of the carrier wave functions, the carriers are predominantly confined to a single layer. This assumption leads to three different bare exciton species: i) intralayer excitons within each individual layer (i.e., ground and excited states of the A and B exciton series in MoSe$_2$ \cite{arora2015exciton}); ii) interlayer excitons with the carriers residing in adjacent layers (IX$_{2L}$), in which the electron and hole occupy the upper spin-orbit-split conduction band and the topmost valence band of each layer, respectively (see Fig. \ref{fig1}b); and iii) interlayer excitons in which the carriers reside in the outermost layers of the 3L MoSe$_2$ region (IX$_{3L}$), where the electron (hole) occupies the lowermost (topmost) conduction (valence) band (see Fig. \ref{fig1}b). In the 3L MoSe$_2$ region, IX$_{2L}$ can be formed with carriers from the middle MoSe$_2$ layer (L$_2$) and carriers from either the bottom (L$_1$) or top (L$_3$) MoSe$_2$ layers. Note that, despite the spatially indirect character of IX$_{2L}$ and IX$_{3L}$, these interlayer exciton species exhibit momentum-direct (intravalley) optical transitions within $\pm$K. Moreover, the vertical displacement of the electron and hole wave functions of the bare interlayer excitons results in an out-of-plane static electric dipole (negligible for intralayer excitons), with a dipole polarity and magnitude that depend on the positions and the spatial separation of the electron and hole in the multilayer, respectively (see Fig. \ref{fig1}a). Such out-of-plane permanent electric dipoles of interlayer excitons in other TMD multilayers, homostructures, and heterostructures have been shown to lead to large shifts of the exciton transition energies via the the quantum confined Stark effect \cite{rivera2016valley,jauregui2019electrical,leisgang2020giant,baek2020highly,lorchat2021excitons,peimyoo2021electrical}. Finally, the natural 2H stacking of our terraced 2L/3L MoSe$_2$ flake results in a different relative stacking configuration between the different layers. While IX$_{2L}$ involves MoSe$_2$ layers with a relative 2H stacking, IX$_{3L}$ originates from MoSe$_2$ layers with a relative R$^h_h$-type stacking, where \emph{h} denotes the hexagon centre of the crystal lattice in each layer (see Fig. \ref{fig1}a). Hence, each IX species is endowed with distinct spin–layer–valley configurations (see Fig. \ref{fig1}b) that can be optically probed. Finally, we note that the relative R$^h_h$-stacking between L$_1$ and L$_3$ is distinct from 3R-stacked homobilayers \cite{paradisanos2020controlling, shi20173r}. For 3R-type bilayers the two layers are laterally shifted so that the hexagon centres in each layer are not vertically aligned, rendering a different symmetry and resulting in a forbidden hole tunneling \cite{gong2013magnetoelectric}. 

\begin{figure*}
    	\begin{center}
    		\includegraphics[scale= 0.18]{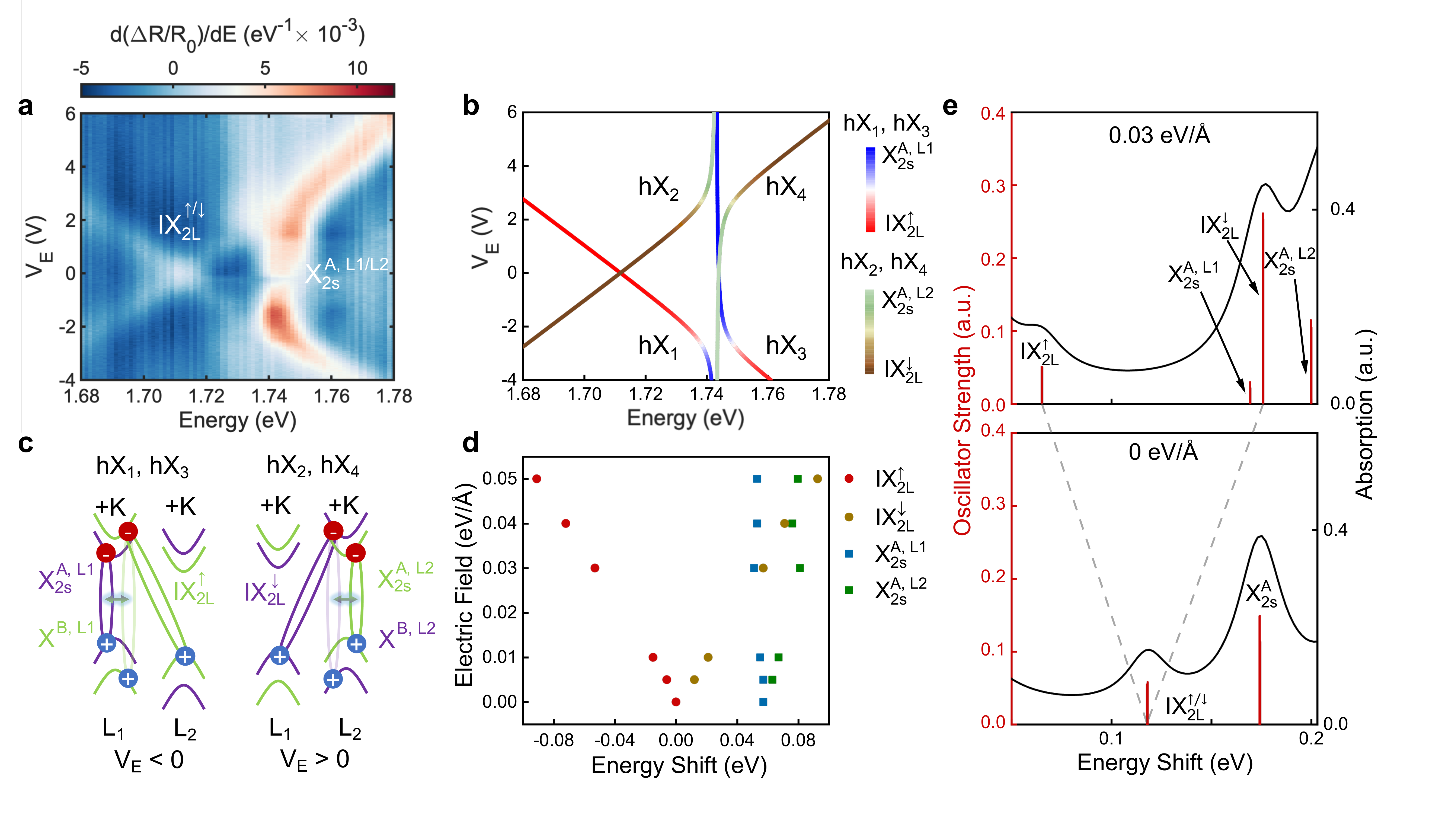}
    	\end{center}

        \caption{\textbf{Layer hybridised excitons in 2L 2H-MoSe$_2$.} \textbf{a}, V$_E$ dependence of the first derivative of the reflectance contrast spectra with respect to photon energy (d($\Delta$R/R$_0$)/dE) in our 2L 2H-MoSe$_2$ in the spectral range 1.68 - 1.78 eV. \textbf{b}, Calculated energies of the different hybrid IX$_{2L}$-X$^{A}_{2s}$ exciton states as a function of V$_E$, which we label as hX$_1$, hX$_2$, hX$_3$, and hX$_4$ from low to high energy at $V_E > 0$ V, respectively. The colour of the solid lines denotes the contribution of the different bare exciton states to each hybrid exciton. \textbf{c}, Schematics of the spin, valley, and layer configuration of the exciton states responsible for the exciton hybridisation shown in panel a for negative and positive applied V$_E$ (left and right panels, respectively). The exciton hybridisation is attributed to a second-order effective coupling between IX$_{2L}$ and the intralayer A exciton facilitated via the A and B exciton admixture (depicted by the glowing double arrows). \textbf{d}, The energy position of IX$_{2L}$ and X$^{A}_{2s}$ as a function of applied electric field, as obtained from $GW$+BSE calculations. The labels identify the simplified bare exciton states. \textbf{e} Normalized theoretical oscillator strengths (red vertical line) and absorption spectra (black line) at 0 eV/\AA~and 0.03 eV/\AA ~focusing on the energy range of IX$_{2L}$ and X$^{A}_{2s}$. The numerical precision of our calculations is estimated to be of the order of $\pm$5 meV, see computational details in Supplementary Note S4.}
        \label{fig2}
\end{figure*}
\subsection{Electric field dependent excitonic transitions in 2L 2H-MoSe$_2$}

To investigate the layer-dependent and intra-/interlayer nature of the different exciton species in the terraced 2H-MoSe$_2$ flake, we perform differential reflection contrast ($\Delta$R/R$_0$) spectroscopy at cryogenic temperature (4 K) as a function of V$_E$ at charge neutrality, where $\Delta$R = R$_s$ - R$_0$, and R$_s$ (R$_0$) is the intensity of the light reflected by the flake (substrate). We focus first on the 2L region of our MoSe$_2$ sample. Supplementary Fig. S1 shows a representative reflection contrast spectrum of our 2L 2H-MoSe$_2$ at $V_E=$ 0 V, where several excitonic resonances with different linewidths and absorption strengths are observed (see Suppl. Fig. S2. for the extracted exciton linewidths). The strongest exciton resonances at low ($\sim$1.63 eV) and high energy ($\sim$1.86 eV) correspond to the ground states of intralayer A (X$^{A}_{1s}$) and B (X$^{B}_{1s}$) neutral excitons in 2L MoSe$_2$, respectively. Two additional weaker exciton resonances, separated only by $\sim$25 meV, are also observed in the energy range between the A and B intralayer 1$s$ exciton states, which can be attributed to IX$_{2L}$ ($\sim$1.715 eV) and the first excited exciton state X$^{A}_{2s}$ of the neutral A exciton ($\sim$1.74 eV), in agreement with previous experimental and theoretical results on bulk \cite{arora2018valley} and 2L 2H-MoSe$_2$ \cite{horng2018observation}. More precisely, we perform $GW$+BSE calculations (see Suppl. Note S4 for computational details) and obtain IX$_{2L}$ and X$^{A}_{2s}$ peaks 0.11 and 0.17 eV above the X$^{A}_{1s}$ one, respectively (see Suppl. Fig. S3), which qualitatively agrees with our experiment in which IX$_{2L}$ and X$^{A}_{2s}$ are 0.07 and 0.11 eV above X$^{A}_{1s}$. To corroborate the intra-/interlayer character of the IX$_{2L}$ and X$^{A}_{2s}$ states and unravel their layer-dependent properties, we show in Fig. \ref{fig2}a a density plot of the V$_E$-dependence of the first derivative of the reflectance contrast spectra with respect to photon energy (d($\Delta$R/R$_0$)/dE) in the spectral range 1.68 - 1.78 eV, which helps to visualise these exciton transitions (see Suppl. Fig. S4 for comparison with bare $\Delta$R/R$_0$ spectra). At V$_E$ = 0 V, we observe the two excitonic resonances attributed to IX$_{2L}$ and X$^{A}_{2s}$. The application of a V$_E$ leads to a stark contrast in the behaviour of the two resonance peaks. For small positive applied V$_E$, the resonance energy of the high energy peak remains almost constant, while the low energy peak splits into two exciton branches which shift symmetrically towards lower and higher energies with a linear dependence with V$_E$. This phenomenon can be interpreted in terms of the DC Stark energy ($\Delta U$) tuning experienced by interlayer excitons under applied out-of-plane electric fields: $\Delta U=-pE$, with $E$ the strength of the vertical electric field and $p=ed$ the out-of-plane electric dipole moment (where $e$ represents the electron charge and $d$ the electron-hole distance). The absolute values of the energy shifts of the two IX$_{2L}$ branches allow us to estimate an average electron-hole spatial separation of $d = 0.34 \pm 0.01$ nm for IX$_{2L}$, which is comparable to the reported values of 0.39-0.47 nm for IXs in other TMD homostructures such as 2L MoS$_2$ \cite{leisgang2020giant,lorchat2021excitons}, and 0.63 nm (0.26 nm) for momentum direct {(indirect)} IXs in twisted bilayer MoSe$_2$ \cite{sung2020broken}. This result unambiguously demonstrates the presence of a sizeable static electric dipole in the out-of-plane direction and corroborates the interlayer exciton nature of IX$_{2L}$ \cite{arora2018valley,horng2018observation}. Moreover, the symmetrical but opposite energy shifts of the two IX$_{2L}$ branches reveal the presence of IX$_{2L}$ excitons with different polarities (i.e., with static electric dipoles aligned parallel and anti-parallel to the applied electric field), as sketched in Fig. \ref{fig1}a. For positive V$_E$, IX$_{2L}$ excitons shifting to lower (higher) energies originate from Coulomb-bound electron-hole pairs in which the hole is spatially located in the top (bottom) layer of the 2L MoSe$_2$, i.e., with static electric dipoles pointing up (IX$_{2L}^\uparrow$) and down (IX$_{2L}^\downarrow$), respectively. Note that for negative V$_E$, the behaviour of IX$_{2L}^{\uparrow(\downarrow)}$ is reversed. Therefore, the Stark shifts of IX$_{2L}^{\uparrow(\downarrow)}$ allow us to unravel their layer configuration.

Further, the large DC Stark tuning of IX$_{2L}$ allows us to explore the possible hybridisation between IX$_{2L}$ and the energetically close X$^{A}_{2s}$ by reducing their energy detuning via the applied V$_E$. For $|V_E|\approx$ 2 V, IX$_{2L}$ and X$^{A}_{2s}$ show an energy anti-crossing characteristic of coupled systems, suggesting the hybridisation of the two exciton species. We note that this observation is reproduced across the 2L 2H-MoSe$_2$ sample.
Supplementary Fig. S5 shows the results for a different spatial location in the 2L 2H-MoSe$_2$. In order to estimate the magnitude of the coupling between IX$_{2L}$ and X$^{A}_{2s}$, we employ a phenomenological model in which the hybridisation between the exciton states is treated as a coupling between oscillators with resonance energies corresponding to the bare exciton states ({see Suppl. Section S3 for details). In our model, we take into account the spin, valley, and layer degrees of freedom of each exciton species, which leads to eight exciton resonances with different spin, valley, and layer properties: two IX$_{2L}$ with opposite polarities (IX$_{2L}^{\uparrow(\downarrow)}$) and momentum-direct transitions at $\pm$K each, and X$^{A}_{2s}$ localised in the top (X$_{2s}^{A,L_2}$) and bottom (X$_{2s}^{A,L_1}$) MoSe$_2$ layers with momentum-direct optical transitions at $\pm$K each. Regarding the IX$_{2L}$-X$^{A}_{2s}$ hybridisation, we include it in our model as a phenomenological interlayer coupling of holes at $\pm$K, with a magnitude which we assume to be independent of V$_E$. In our calculations, the energies of the exciton states at $V_E=$ 0 V and the slope of the DC Stark shift for IX$_{2L}^{\uparrow(\downarrow)}$ are set to match the corresponding experimental values, while the value of the coupling strength between the exciton states is left as a free parameter that can be tuned to fit our experimental data. Figure \ref{fig2}b shows the calculated energies of the resulting hybrid exciton states as a function of V$_E$, which we label as hX$_1$, hX$_2$, hX$_3$, and hX$_4$ from low to high energy at V$_E>0$ V, respectively. The colour of the solid lines in Fig. \ref{fig2}b denotes the contribution of the different bare exciton states to each hybrid exciton. The phenomenological model captures well the hybridisation-induced renormalisation of the exciton resonance energies with increasing electric field, allowing us to estimate a IX$_{2L}$-X$^{A}_{2s}$ coupling strength $\kappa_{2L-2s}\approx$ 5.2 meV, which is slightly smaller but of the same order of magnitude as the linewidths of the exciton states at $V_E=0$ V (see Suppl. Fig. S2).

The physical origin of the observed hybridisation between an interlayer exciton and the first excited state of an intralayer exciton is intriguing, and to the best of our knowledge has not been been reported in any other homobilayer TMD system. We discard the possibility of spin-conserving electron hopping between the electron states involved in IX$_{2L}$ and X$^{A}_{2s}$, since such interlayer electron hopping is forbidden in 2L 2H-MoSe$_2$ due to the $C_3$ symmetry of the $d_{z^2}$ orbitals of the conduction band states at $\pm$K \cite{gong2013magnetoelectric,hagel2021exciton}. However, a recent theoretical work has shown that the application of a vertical electric field can lead to the hybridisation of IX$_{2L}$ and X$_{1s}^B$ in 2L 2H-MoSe$_2$ via spin-conserving interlayer hole tunneling \cite{hagel2022electrical}, in agreement with previous experimental and theoretical results for 2L 2H-MoS$_2$ \cite{leisgang2020giant,PhysRevLett.129.107401}. Moreover, similar to 2L 2H-MoS$_2$, the theoretical results in Ref. \cite{hagel2022electrical} also suggest a weak admixture between the A and B excitons in 2L 2H-MoSe$_2$, which leads to a non-vanishing second-order effective hybridisation of IX$_{2L}$ and the intralayer X$^{A}_{1s}$ exciton. Since X$^{A}_{1s}$ and X$^{A}_{2s}$ have the same nature (i.e., same exact spin and valley configuration), we tentatively attribute the observed IX$_{2L}$-X$^{A}_{2s}$ hybridisation to a second-order effective coupling between IX$_{2L}$ and the intralayer A exciton facilitated via the A and B exciton admixture (see Fig. \ref{fig2}c) \cite{PhysRevLett.129.107401,hagel2022electrical}. This result agrees well conceptually with the theoretical modeling in Ref. \cite{hagel2022electrical}, and leads to a layer selective coupling between IX$_{2L}$ and the intralayer A exciton, as also captured by our phenomenological model (see Fig. \ref{fig2}b). Our estimated value of the coupling strength $\kappa_{2L-2s}\approx$ 5.2 meV is very similar to the coupling strength reported for IX$_{2L}$ and X$^{A}_{1s}$ excitons in 2H-MoS$_2$, and significantly smaller than the direct coupling between the IX$_{2L}$ and X$_{1s}^B$ excitons in the same 2H-MoS$_2$ sample \cite{leisgang2020giant,PhysRevLett.129.107401}, which supports our hypothesis.

To further corroborate our hypothesis, we calculate $GW$+BSE-based estimates of the excitonic transition energies and their corresponding oscillator strengths at various external electric field values (see Figs. \ref{fig2}d to \ref{fig2}e). For vertical applied positive electric fields, the IX$_{2L}$ excitons experience a significant Stark shift, which supports their interlayer nature. The X$^{A,L{_1}}_{2s}$ remains relatively unchanged, while the X$^{A,L_{2}}_{2s}$ exhibits an energy shift with the applied electric field, indicative of the layer-selective mixing with the IX$_{2L}$ state. In fact, our calculations reveal a sizeable mixing between IX$_{2L}$ and X$^{A,L_{2}}_{2s}$ even at zero applied electric field. Table 1 in Suppl. Section S4 summarises the oscillator strengths of the different bare optical transitions contributing to each excitonic resonance in Figs. \ref{fig2}d and \ref{fig2}e, corresponding to 0 and 0.03 V/{\AA} applied electric fields, respectively.

\begin{figure*}
    \begin{center}
    \includegraphics[scale= 0.13]{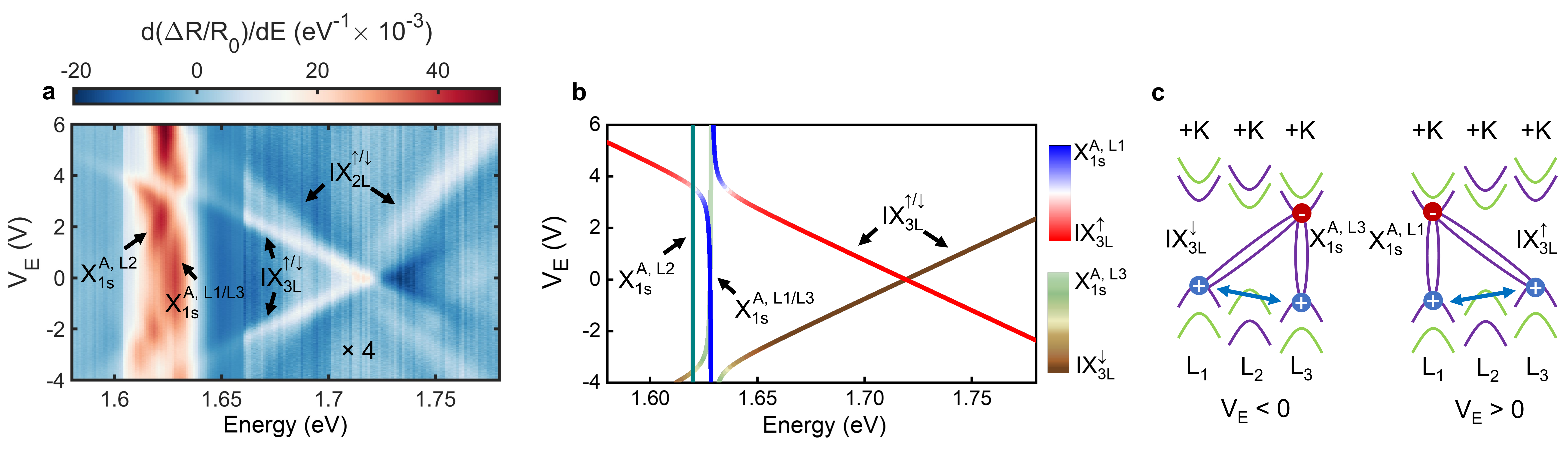}
    \end{center}
    \caption{\textbf{Layer hybridised excitons in 3L MoSe$_2$}. \textbf{a},
    V$_E$ dependence of d($\Delta$R/R$_0$)/dE in the 3L MoSe$_2$ region of our sample. \textbf{b}, Energies of the different hybrid IX$_{3L}$-X$^{A}_{1s}$ exciton states as a function of V$_E$, where the colour of the solid lines denotes the contribution of the different bare exciton states to each hybrid exciton. \textbf{c}, Schematics of the spin, valley, and layer configuration of the exciton states responsible for the exciton hybridisation shown in panel a for negative and positive applied V$_E$ (left and right panels, respectively). The exciton hybridisation is attributed to direct spin-conserving interlayer hole tunnelling between L$_1$ and L$_3$. }
    \label{fig3}
\end{figure*}

\subsection{Electric field dependent excitonic transitions in 3L 2H-MoSe$_2$}


To explore the potential for multilayer TMDs to host IX with dipole moments even larger than IX$_{2L}$ and with greater tunability, we optically probe the 3L region of our terraced 2H-MoSe$_2$ sample at charge neutrality as a function of V$_E$. Figure \ref{fig3}a shows the V$_E$-dependence of d($\Delta$R/R$_0$)/dE in the spectral range 1.58 - 1.78 eV (with the values in the energy range 1.66 - 1.78 eV multiplied by a factor 4 for visualisation purposes). The reflectance spectrum at V$_E$ = 0 V is markedly different to the one observed in the 2L region (see Fig. \ref{fig2}a). We observe two exciton transitions in the energy range corresponding to X$^{A}_{1s}$ with an energy splitting of $\sim$11 meV, and two exciton resonances in the energy range of IX$_{2L}$ with an energy splitting of $\sim$ 14.5 meV. The resonances at low energy can be attributed to X$^{A}_{1s}$ excitons localised in the different layers of our sample, in which the lower average permittivity environment of L$_1$ and L$_3$ (h-BN/L$_{1,(3)}$/MoSe$_2$) compared to L$_2$ (MoSe$_2$/L$_2$/MoSe$_2$) results in a dielectric-induced energy blue shift for X$^{A}_{1s}$ excitons in L$_1$ and L$_3$ (X$_{1s}^{A,L_1(L_3)}$) compared to X$^{A}_{1s}$ excitons in L$_2$ (X$_{1s}^{A,L_2}$) \cite{raja2017coulomb}, similar to what has been observed for 3L MoS$_2$ \cite{leisgang2020giant}. 


To unravel the nature of the excitons in the energy range corresponding to IX$_{2L}$, we focus on their behaviour as a function of V$_E$. As shown in Fig. \ref{fig3}a, the application of a vertical electric field gives rise to a Stark-effect-induced splitting of each resonance into two distinct exciton branches which shift symmetrically towards lower and higher energies with a linear dependence with V$_E$ (see comparison between $\Delta$R/R$_0$ and d$(\Delta$R/R$_0$)/d$E$ in Suppl. Fig. S6). For the lower exciton resonance, we estimate an average electron-hole spatial separation of $d = 0.36 \pm0.01$ nm, which agrees very well with the value found for IX in the 2L region of our sample, confirming its IX$_{2L}$ nature. The energy splitting of the two branches belonging to the higher energy resonance yields an average electron-hole separation of $d = 0.73 \pm0.01$ nm, which exceeds the interlayer distance of TMD homobilayers \cite{liu2014evolution} and is approximately twice the IX$_{2L}$ dipole size. We therefore identify this resonance as IX$_{3L}$. Furthermore, the larger Stark shift of IX$_{3L}$ allows us to tune their energy in resonance to X$^{A}_{1s}$ (see Fig. \ref{fig3}a). For $|V_E|\sim$ 4 V we observe a clear avoided crossing between IX$_{3L}$ and X$_{1s}^{A,L_1/L_3}$, indicative of the hybridisation between the exciton species with carriers hosted in the outermost layers of the 3L MoSe$_2$ sample. Supplementary Fig. S7 shows the results for a different spatial location in the 3L 2H-MoSe$_2$. We note that, contrary to the results for the 2L region, we observe negligible oscillator strength of X$^{A}_{2s}$ in the 3L MoSe$_2$ region, which prevents clear observation of coupling between this exciton state and the two IX species. Therefore, we focus on the clear IX$_{3L}$-X$^{A}_{1s}$ hybridisation. 

Similar to the 2L case, we simulate the V$_E$-dependent energy dispersion of the hybridised excitons in the 3L MoSe$_2$ system using a phenomenological model of coupled oscillators, in which the exciton coupling is both spin- and layer-selective (see Fig. \ref{fig3}b). In this case, we include three different X$^{A}_{1s}$ excitons (i.e., one in each layer) and two IX$_{3L}$ with opposite polarities (IX$_{3L}^{\uparrow(\downarrow)}$) and momentum-direct transitions at $\pm$K each. Spin conserving hole tunneling results in a layer-selective coupling between IX$_{3L}$ and X$^{A}_{1s}$; the polarity of IX$_{3L}$ is locked to the layer degree of freedom of X$_{1s}^{A}$ (e.g.  IX$_{3L}^{\uparrow(\downarrow)}$ only couples to X$_{1s}^{A,L_1(L_3)}$). However, the energy degeneracy of X$_{1s}^{A,L_1}$ and X$_{1s}^{A,L_3}$ prevents clear observation of such layer-selective coupling experimentally. Figure \ref{fig3}b shows the results of the best fit of the model to our experimental data. The colour of the solid lines denotes the contribution of the different bare exciton states to each hybrid exciton. Overall, the phenomenological model captures well the hybridisation-induced renormalisation of the exciton resonance energies with increasing electric field, allowing us to estimate a IX$_{3L}$-X$^{A}_{1s}$ coupling strength $\kappa_{3L-1s}\approx$ 7.5 meV, which is slightly larger but of the same order of magnitude as $\kappa_{2L-2s}$ in the 2L region. Furthermore, it is worth noting that the R-type relative stacking of the layers $L_1$ and $L_3$ leads to some differences between the IX$_{3L}$-X$^{A}_{1s}$ and IX$_{2L}$-X$^{A}_{2s}$ couplings. The $R$-type stacking between $L_1$ and $L_3$ results in IX$_{3L}$ with electron-hole pairs with the same spin-valley configurations as X$^A$ (see Fig. \ref{fig1}a), which allows the hybridisation of the two exciton species via direct spin-conserving interlayer hole tunnelling in the presence of an additional layer between $L_1$ and $L_3$ \cite{shimazaki2020strongly,zheng2022anomalous}. For the specific stacking registry of $R^h_h$ between $L_1$ and $L_3$, a recent work calculated an 11 meV tunnel splitting of the valence band (with monolayer BN as $L_2$) \cite{shimazaki2020strongly}, supporting our interpretation. On the contrary, due to the 2$H$ relative stacking between adjacent layers, the direct hole tunneling between the valence band states in IX$_{2L}$ and X$^A$ has to compete with a sizable detuning equal to the spin splitting at the valence band edges (150 meV in MoSe$_2$ \cite{zheng2022anomalous}), and is thus only facilitated by the admixture of X$^A$ and X$^B$. Finally, we note that the energy detuning between IX$_{2L}$ and the energy degenerate X$_{1s}^{A,L_1}$ and X$_{1s}^{A,L_3}$ exciton resonances is slightly smaller in the 3L region as compared to the 2L one. As a consequence, we are able to observe experimental signatures of the coupling between these two exciton species at $V_E\sim 6$ V (see Fig. \ref{fig3}a and Suppl. Section S6), which corroborate our results for the 2L 2H-MoSe$_2$.

\begin{figure*}
    \begin{center}
    \includegraphics[scale= 0.17]{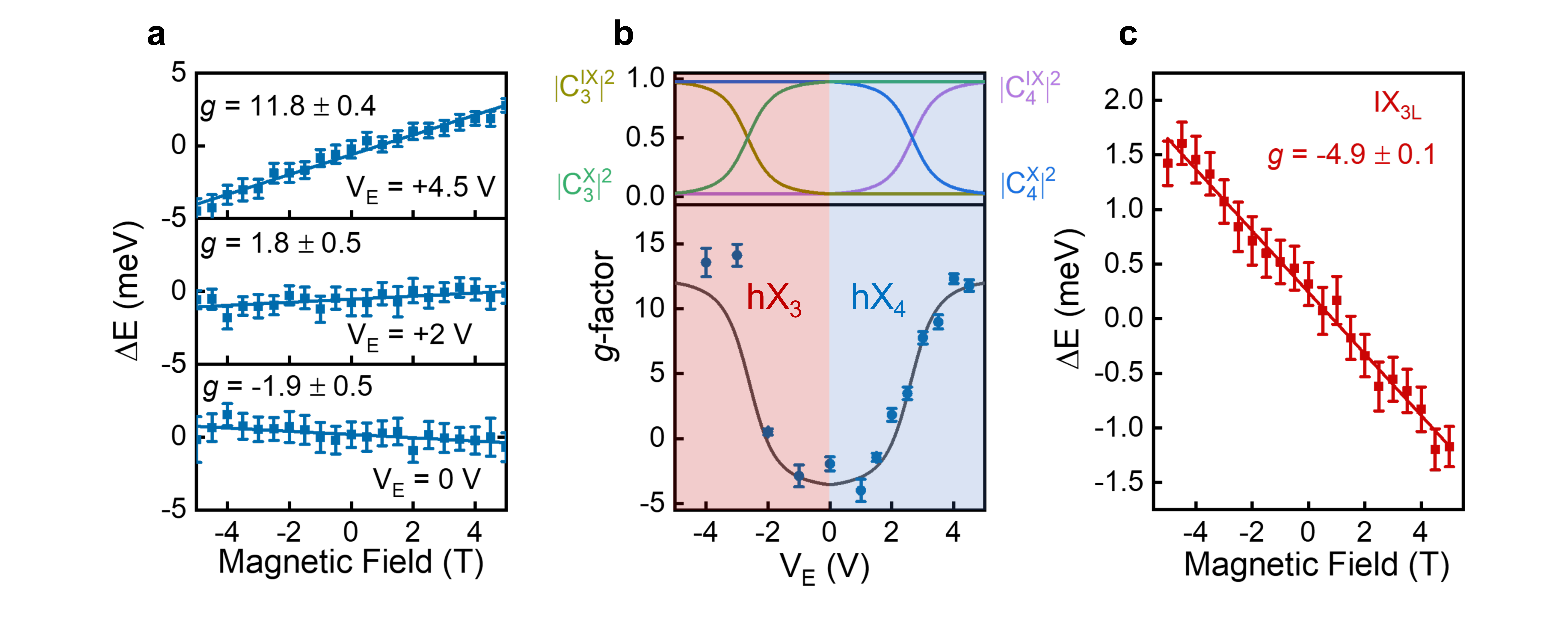}
    \end{center}
    \caption{\textbf{Magneto-optical properties of layer-hybridised excitons in 2L MoSe$_2$ and 3L MoSe$_2$}. \textbf{a}, Zeeman splitting of hX$_4$ at three different applied V$_E$. The blue dots represent the experimental values, while the blue solid lines show linear fits of the experimental data, from which we are able to estimate the effective $g$-factor of this hybrid exciton at each applied V$_E$. \textbf{b}, V$_E$-driven evolution of the $g$-factor of the hybrid excitons hX$_3$ (-5 to 0 V, red shaded area) and hX$_4$ (0 to 5 V, blue shaded area) in bilayer MoSe$_2$ (bottom panel). The top panel shows the V$_E$-dependent contributions of each bare exciton state $|C_{3,4}^{IX(X)}|^2$ to the corresponding hybrid excitons. \textbf{c}, Zeeman splitting of IX$_{3L}$ at measured at $V_E=0$ V.}
    \label{fig4}
\end{figure*}

\subsection{Electric-field-dependent magneto-optical properties of hybrid excitons in MoSe$_2$}


Beyond tuning of the exciton resonance energies, oscillator strengths, and effective permanent electric dipoles, here we explore if the electric-field-dependent control of the exciton nature enables precision tuning of the effective Land{\'e} $g$-factors. The application of a vertical magnetic field $B$ results in the Zeeman splitting of the optical transitions of each exciton at $\pm$K, with an energy splitting $\Delta E(B) = g\mu_0B$, with $\mu_0$ the Bohr magneton. Optical transitions at $\pm$K can be independently probed by $\sigma^{\pm}$-polarised light, respectively, which enables us to perform circularly polarised reflectance contrast measurements to estimate the experimental Zeeman splitting $\Delta E=E^{\sigma^+}-E^{\sigma^-}$, with $E^{\sigma^{\pm}}$ the energy of the $\sigma^{\pm}$-polarised transition. We focus first on the hybrid IX$_{2L}$-X$^{A}_{2s}$ exciton states in our 2L region. Figure \ref{fig4}a shows the measured Zeeman splittings (blue dots) for hX$_4$ at three different applied V$_E$ values. Supplementary Figure S8 shows the linecuts for the $\sigma^{\pm}$-resolved d($\Delta$R/R$_0$)/d$E$ at $V_E=0$~V for 5~T. The blue solid lines represent linear fits of the experimental data, from which we estimate the effective $g$-factor at each applied V$_E$. 
We observe the effective $g$-factor of hX$_4$ is tuned from a negative value ($-1.9\pm0.5$) to a relatively large positive value ($11.8\pm0.4$). To explore this effect in more detail, we employ the same experimental procedure and extract the effective $g$-factors of hX$_3$ and hX$_4$ in the range of applied V$_E$ in which the oscillator strength of each transition and their energy detuning with respect to other transitions enable a reliable estimate of $E^{\sigma^{\pm}}$, i.e., $V_E<0$ for hX$_3$ and $V_E>0$ for hX$_4$ (as indicated by the red and blue shaded areas in Fig. \ref{fig4}b). The results highlight a continuous and smooth transition of the $g$-factor of the two hybrid exciton states from $-1.9\pm0.5$ to $11.8\pm0.4$ by changing V$_E$ from 0 V to $\pm4.5$ V. Such evolution of the effective $g$-factor arises from the V$_E$-dependent hybridisation between IX$_{2L}$ and X$^{A}_{2s}$, and can be quantitatively explained by our phenomenological model of coupled oscillators. In this model, the wave function of each hybrid exciton is expressed as a superposition of the bare intralayer and interlayer exciton wave functions (see Suppl. Section S3). In the case of hX$_3$ and hX$_4$, these exciton wave functions are expressed as follows:
\begin{align}
    	\ket{hX_3(V_E)}=C_{3}^{IX}(V_E)\ket{IX_{2L}^{\uparrow}}+C_{3}^{X}(V_E)\ket{X_{2s}^{A,L_1}}
    \label{Eq:state_projection}
\end{align}
and
\begin{align}
    	\ket{hX_4(V_E)}=C_{4}^{IX}(V_E)\ket{IX_{2L}^{\downarrow}}+C_{4}^{X}(V_E)\ket{X_{2s}^{A,L_2}},
    \label{Eq:state_projection}
\end{align}
where $C_i^{IX(X)}(V_E)$ represents the V$_E$-dependent amplitude of the bare interlayer (intralayer) exciton state in hybrid exciton state $\ket{hX_i}$, with $|C_i^{IX}(V_E)|^2+|C_i^{X}(V_E)|^2=1$. Therefore, the effective $g$-factor of each hybrid exciton state ($g_{hX_i}$) can be expressed as
\begin{align}
    	g_{hX_i}(V_E)=|C_i^{IX}(V_E)|^2g_{IX_{2L}}+|C_i^{X}(V_E)|^2g_{X_{2s}},
    \label{Eq:g_factor}
\end{align}
where $g_{IX_{2L}(X_{2s})}$ represents the $g$-factor of the bare IX$_{2L}$ (X$_{2s}$) state. The solid line in Fig. \ref{fig4}b represents a fit of the experimental data to Eq. (\ref{Eq:g_factor}), in which we have used the $C_i^{IX(X)}(V_E)$ values shown in the top panel of Fig. \ref{fig4}b (obtained from the fits in Fig. \ref{fig2}b), and $g_{IX_{2L}}$ and $g_{X_{2s}}$ have been left as free fitting parameters. Our model captures well the hybridisation-induced evolution of the $g$-factors and allows us to estimate of the effective $g$-factors of the bare states: $g_{X_{2s}}\approx-4$ and $g_{IX_{2L}}\approx12.5$. We note that, although to the best of our knowledge $g_{X_{2s}}$ has not been previously reported for 2L MoSe$_2$, our estimated value of $g_{X_{2s}}$ is in very good agreement with reported experimental ($-3.6\pm0.1$ \cite{arora2018zeeman}) and theoretical (-3.7 \cite{deilmann2020ab}) values for this excited exciton state in monolayer MoSe$_2$ and other TMD systems such as bulk WSe$_2$ ($-3.3\pm0.6$ \cite{arora2018valley}). Regarding $g_{IX_{2L}}$, the estimated sign and value are also in good agreement with the value predicted by a simplistic “atomic picture”, in which the $g$-factor of the bands hosting the electron-hole pairs are assumed to be equal to the sum of their spin, orbital, and valley magnetic moments \cite{aivazian2015magnetic}. Within this model, the spin-valley configuration of IX$_{2L}$ results in a $g_{IX_{2L}}=2(\frac{m_0}{m^{*}_{c}}+\frac{m_0}{m^{*}_{v}})+4\approx 9.7$, with m$_0$ the free electron mass, and m$^{*}_{c}$ and m$^{*}_{v}$ the experimentally reported electron and hole effective masses for the bottom conduction band and top valence band in monolayer MoSe$_2$, respectively \cite{goryca2019revealing}. The relatively large and positive value of the $g$-factor estimated with this atomic picture model provides an additional confirmation of the spin-valley configuration and interlayer nature of IX$_{2L}$. Also, we note that the small discrepancy between the experimental and the calculated value for $g_{IX_{2L}}$ might arise from a combination of several factors, including the limitation of this simple model to estimate accurately the $g$-factor of the relevant bands \cite{arora2021magneto} and the absence of experimental values for the electron effective mass of the top conduction band in 2L MoSe$_2$, which theoretical calculations predict to be slightly smaller than for the bottom conduction band in Mo-based TMDs such as 2H-MoSe$_2$ \cite{kormanyos2015k} and 2H-MoS$_2$ \cite{brotons2018optical}. Nevertheless, we note that the extracted $g$-factors of the bare IX$_{2L}$ and X$_{2s}$ states allow us to include the effects of the Zeeman splitting in the calculated V$_E$-dependent evolution of the hybrid IX$_{2L}$-X$_{2s}$ states, for which we find a very good agreement with the experimental results (see Suppl. Note S5 and Suppl. Fig. S12).

Finally, we investigate the magneto-optical properties of IX$_{3L}$ in the 3L region of the sample. Figure \ref{fig4}c shows the experimental Zeeman splitting for this exciton species at $V_E=0$ V, from which we estimate a $g$-factor $g_{IX_{3L}} = -4.9\pm 0.1$. The extracted value for $g_{IX_{3L}}$ is very similar to the experimentally reported $g$-factor of X$^{A}_{1s}$ in 3L MoSe$_2$ \cite{arora2021magneto}, which confirms the identical spin-valley configuration of these two exciton species and corroborates that the giant dipole IX$_{3L}$ indeed originates from electron and hole in L$_{1}$ and L$_{3}$ (or vice versa), respectively. The identical spin-valley configuration of X$^{A}_{1s}$ and IX$_{3L}$ has an important consequence on the magneto-optical properties of the hybrid X$^{A}_{1s}$-IX$_{3L}$ states: contrary to the hybrid X$^{A}_{2s}$-IX$_{2L}$ states in 2L MoSe$_2$, hybrid X$^{A}_{1s}$-IX$_{3L}$ do not feature a V$_E$-tunable $g$-factor, since the two bare exciton states already exhibit a similar $g$-factor.

    \subsection{Excited state interlayer excitons in multilayer 2H-MoSe$_2$}
    
    Although excited (Rydberg) exciton states have weaker oscillator strength than their corresponding ground states \cite{chernikov2014exciton}, in this section we show that the hybridisation of interlayer excitons with intralayer transitions leads to clear spectroscopic signatures of excited Rydberg states for both IX$_{2L}$ and IX$_{3L}$. Figures \ref{fig5}a and \ref{fig5}b show d$^2$($\Delta$R/R$_0$)/d$E^2$ spectra as a function of V$_E$ for 2L and 3L 2H-MoSe$_2$ regions, respectively, corresponding to spatial locations where the interlayer excited exciton states present appreciable oscillator strengths. The data shown in the the spectral ranges 1.797 - 1.830 eV and 1.665 - 1.800 eV in Figs. \ref{fig5}a and \ref{fig5}b have been multiplied by a factor 2 for visualisation purposes. In the 2L region (Fig. \ref{fig5}a), for $4\lesssim V_E \lesssim 6$ V we resolve an additional exciton transition in the energy range $\sim$1.7 eV with a linear Stark shift parallel to the IX$_{2L}^{\uparrow}$. Extrapolation of the linear energy shift to $V_E=0$ V gives an estimated energy of $\sim$ 1.77 eV for this excitonic transition with clear interlayer nature. We note that the relative energy position of this excitonic peak with respect to both IX$_{2L}$ and X$^{A}_{2s}$ agrees well with the first excited Rydberg state of IX$_{2L}$ predicted by our $GW-$BSE results (see Figs. \ref{fig2}e and Suppl. Fig. S3). This, together with the dipole moment of 0.30 $\pm$ 0.01 enm estimated from the linear Stark shift (almost identical to the one for IX$_{2L}$), allows us to attribute the observed resonance to the first excited Rydberg state of IX$_{2L}^{\uparrow}$ (i.e., IX$_{2s,2L}^{\uparrow}$). Notably, for $|V_E|\sim3$ V and energies $\sim$ 1.82 eV, we observe two additional resonances with opposite Stark shift slopes that also extrapolate to an energy of $\sim$1.77 eV at $V_E=0$ V, which we attribute to IX$^{\downarrow}_{2s, 2L}$ ($V_E>0$ V) and IX$^{\uparrow}_{2s, 2L}$ ($V_E<0$ V), corroborating our peak assignment. We note that the IX$_{2s,2L}$ state is a general feature rather than a location dependent feature of our sample, as we also observe IX$_{2s, 2L}$ in the main locations of 2L and 3L MoSe${_2}$ shown in Figs.  \ref{fig2}-\ref{fig4}, as depicted in Suppl. Fig. S10. Further, we observe a very weak transition near 1.72 eV for $4\lesssim V_E \lesssim 6$ V with a Stark shift parallel to IX$_{2s,2L}$ (see Suppl. Fig. S9), which we tentatively ascribe as the $3s$ Rydberg state of IX$_{2L}$. Extrapolation to V$_E$ = 0 V gives an estimated energy for the $3s$ state of $\sim$20 meV higher than the $2s$ at zero applied electric field. 
    
    Similar to Fig. \ref{fig5}a, the results in Fig. \ref{fig5}b also display clear excited Rydberg states in our 3L 2H-MoSe$_2$ sample. In addition to the coexisting interlayer exciton species IX$_{2L}$ and IX$_{3L}$ already shown in Fig. \ref{fig3}, we observe two additional transitions with Stark-induced linear energy shifts parallel to the IX$_{3L}$, which we attribute to the two opposite polarities of the $2s$ excited Rydberg state  of IX$_{3L}$ (i.e., IX$^{\uparrow(\downarrow)}_{2s, 3L}$). Extrapolation of the measured linear energy shifts of IX$^{\uparrow(\downarrow)}_{2s, 3L}$ to V$_E=0$ V gives an estimated energy $\sim$ 40 meV above the IX$_{3L}$ ground state (near the X$^{A}_{2s}$ transition).

\begin{figure*}
    \begin{center}
    \includegraphics[scale= 0.18]{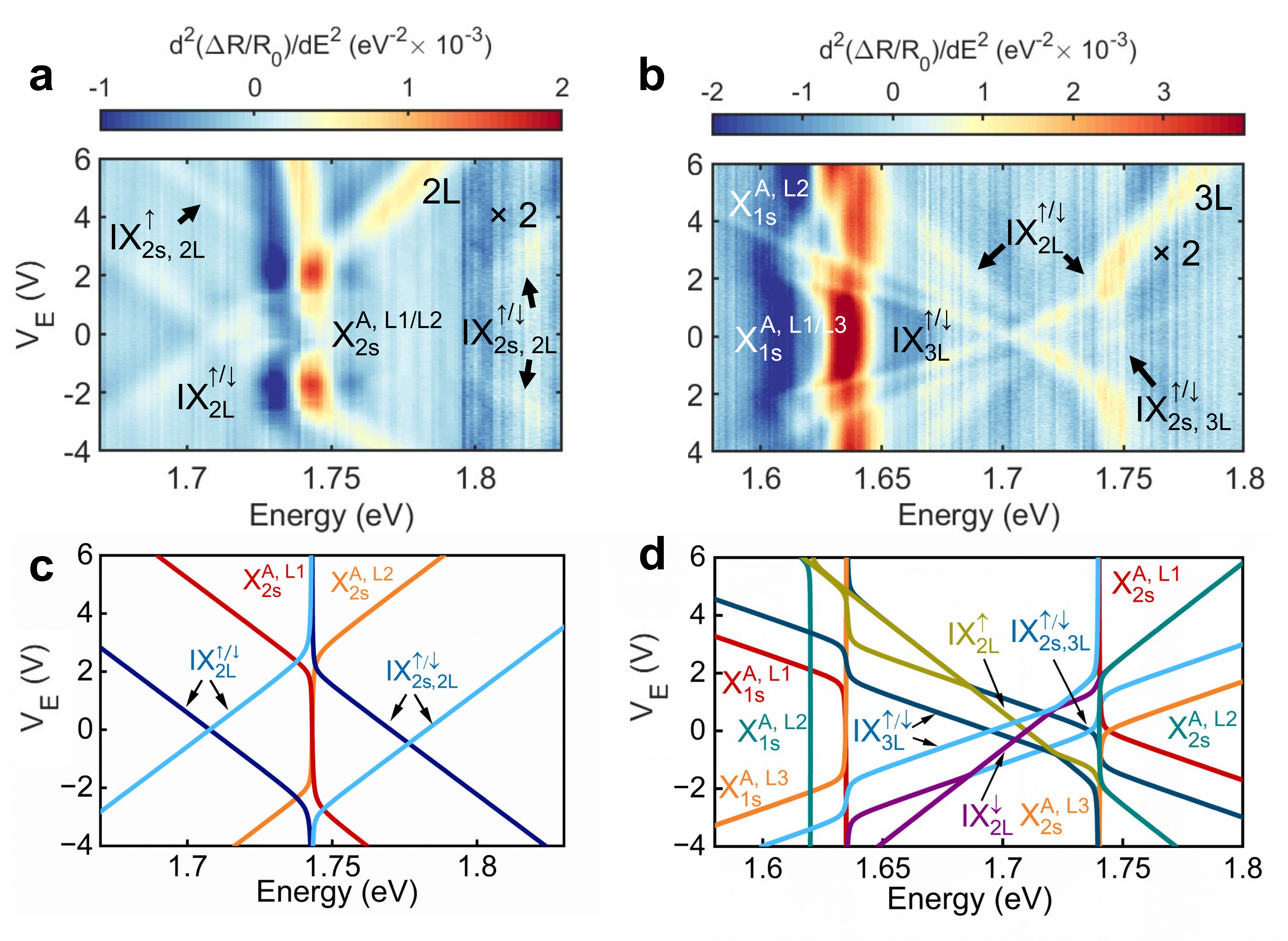}
    \end{center}
    \caption{\textbf{Observation of excited states of IX$_{2L}$ and IX$_{3L}$.} \textbf{a}, $V_E$ dependence of d${^2}$($\Delta$R/R${_0}$)/dE$^{2}$ in a second location of the 2L 2H-MoSe${_2}$ sample in the spectral range 1.67 - 1.83 eV. \textbf{b}, d$^2$($\Delta$R/R${_0}$)/dE$^{2}$ in another location of the 3L 2H-MoSe${_2}$ sample in the spectral range 1.58 - 1.8 eV. \textbf{c} Calculated energies of the different exciton states of 2L 2H-MoSe${_2}$ including IX$_{2s, 2L}$ with same spin and layer configuration as IX$_{2L}$. \textbf{d} Calculated energies of the different exciton states in 3L 2H-MoSe$_2$ including IX$_{2s, 3L}$ with the same spin and layer configuration as IX$_{3L}$.}
    \label{fig5}
\end{figure*}

Finally, we simulate the V$_E$-dependent energy dispersions of the hybridised ground and excited exciton Rydberg states using the phenomenological coupled oscillator model previously described. We build on the models used for Figs. \ref{fig2}b and \ref{fig3}b and add the the observed excited Rydberg states, which we assume to have identical Stark shifts and the same spin- and layer-selective couplings than their corresponding ground states. Figures \ref{fig5}c and \ref{fig5}d show the simulated energy dispersions corresponding to the results in Figs. \ref{fig5}a and \ref{fig5}b, respectively. As can be observed in these figures, the simulated energy dispersions capture well the $V_E$-induced hybridisation and energy dispersion of the different excitonic transitions (see Suppl. Fig. S15 for the the simulated absorption spectrum as a function of V$_E$ corresponding to Figs. \ref{fig5}b using this model and suppl. section 6 for simulation details), supporting our interpretation of their different intra-/interlayer origin, ground/excited state character, and spin and valley configurations.

\section{Summary and Outlook}
Our work reports the observation of giant Stark splitting of the interlayer excitons in 2L and 3L 2H-MoSe$_2$. First, we observe hybridisation between IX$_{2L}$ and X$^{A}_{2s}$. In addition to their spectral evolution, a hybridisation driven g-factor evolution of the coupled excitons is resolved and understood via a unified coupled oscillator model. The ability to drive the exciton Zeeman splitting from negative to positive through zero has potential in electrically tunable valleytronics and spin-dependent exciton-exciton interactions. Next, we report the giant excitonic trilayer dipole IX$_{3L}$, which exhibits distinct spin-layer selection rules for hybridisation with X$^{A}_{1s}$.  A salient feature of the large IX$_{3L}$ dipole is that it can be Stark tuned to become the ground state (e.g. lower energy than X$^{A}_{1s}$), promising for applications in exciton transport.  Finally, by harnessing the exciton hybridisation affects, we successfully probe the excited state IX for the first time in a TMD system. Future theory and experimental effort is encouraged to better understand the excited IX states, including comparisons with the hydrogen model and the magnetic field and doping dependence to reveal the g-factor, effective mass \cite{goryca2019revealing} and Roton-like \cite{liu2021exciton} properties of the interlayer Rydberg excitons. Altogether, the results promote a new TMD homostructure candidate for applications with enhanced exciton-exciton interactions with strong light-matter coupling. Beyond 3L 2H-MoSe$_2$, a strategy of further engineering IX dipoles by tuning the layer number, including thicker multilayer (> 3L) 2H-TMDs or heterostrutures with multilayer TMD components and hBN spacers, is encouraged.

    

\section{Acknowledgements}

S.F. and A.C. contributed equally to this work. This work was supported by the EPSRC (grant nos. EP/P029892/1 and EP/L015110/1), the ERC (grant no. 725920). S.F. is supported by a Marie Skłodowska-Curie Individual Fellowship H2020-MSCA-IF-2020 SingExTr (No. 101031596). M.B.-G. is supported by a Royal Society University Research Fellowship. B.D.G. is supported by a Wolfson Merit Award from the Royal Society and a Chair in Emerging Technology from the Royal Academy of Engineering. K.W. and T.T. acknowledge support from the Elemental Strategy Initiative conducted by the MEXT, Japan (Grant Number JPMXP0112101001) and JSPS KAKENHI (Grant Numbers 19H05790, 20H00354 and 21H05233). I.C.G acknowledges the CALMIP initiative for the generous allocation of computational time, through Project No. p0812, as well as GENCI-CINES, GENCI-IDRIS, GENCI-CCRT for Grant No. A012096649.

\appendix*

\bibliographystyle{apsrev4-1}
\bibliography{Manuscript}

\end{document}